\documentclass[aps,pra,reprint,groupedaddress]{revtex4-1}
\pdfoutput=1
\usepackage[none]{hyphenat}
\usepackage{graphicx}
\usepackage{multirow}
\usepackage{mathrsfs}
\usepackage{amsfonts}
\usepackage{amssymb}
\usepackage{cancel}
\usepackage{amsmath}
\begin{document}
\title{A gradient algorithm for Hamiltonian identification of open quantum systems}
\author{Shibei Xue$^{1,2}$}\email[]{shbxue@sjtu.edu.cn} \author{Rebing Wu $^{3,4}$} \author{Dewei Li$^{1,2}$} \author{Min Jiang$^{2,5}$}
\affiliation{$^1$Department of Automation, Shanghai Jiao Tong University, Shanghai 200240, P. R. China.}
\affiliation{$^2$Key Laboratory of System Control and Information Processing, Ministry of Education of China, Shanghai 200240, P. R. China.}
\affiliation{$^3$Department of Automation, Tsinghua University, Beijing 100084, P. R. China.}
\affiliation{$^4$Center for Quantum Information Science and Technology, BNRist, Beijing 100084, P. R. China.}
\affiliation{$^5$School of Electronics and Information Engineering, Soochow University, Suzhou 215006, P. R. China.}



\date{\today}

\begin{abstract}
In this paper, we present a gradient algorithm for identifying unknown parameters in an open quantum system from the measurements of time traces of local observables. The open system dynamics is described by a general Markovian master equation based on which the Hamiltonian identification problem can be formulated as minimizing the distance between the real time traces of the observables and those predicted by the master equation. The unknown parameters can then be learned with a gradient descent algorithm from the measurement data. We verify the effectiveness of our algorithm in a circuit QED system described by a Jaynes-Cumming model whose Hamiltonian identification has been rarely considered. We also show that our gradient algorithm can learn the spectrum of a non-Markovian environment based on an augmented system model.
\end{abstract}

\pacs{}

\maketitle

\section{Introduction}
Quantum information technology has attracted much attention in the past decades, including physically absolutely secure  quantum communication and huge accelaration in quantum computation~\cite{Nielsen}. These advantages rely on high precision quantum operations designed with good models of the quantum information carriers, e.g., superconducting qubits or quantum dots, etc.
However, it would not be easy to obtain a complete model for a quantum system. For example, in a recent experiment on a quantum dot system, the calculation based on a Markovian system model without containing the exact effect of noises arising in the environment fails to match the experiment data~\cite{Frey}. In addition, the inaccuracy of parameters in a model of quantum systems would also lead to errors in quantum computation.

Quantum identification is the crucial step towards obtaining a complete and accurate model, which utilizes the  measurement data of excited quantum dynamics to extract unknown information. In particular, Hamiltonian identification has gradually become a hot topic, whose task is to estimate unknown parameters in the Hamiltonian or to identify unknown structures in a quantum system.
For closed quantum systems,
an inversion-algorithm-based learning approach was presented to identify both the free hamiltonian and the dipole for a class of molecular systems~\cite{PhysRevLett.89.263902}. Employing the system's algebraic properties, the identification algorithms were proposed for special finite-level quantum systems, such as two-level systems~\cite{PhysRevA.69.050306,ISI000278904200017}, spin-1 systems~\cite{ISI000230468800015} and spin networks~\cite{ALBERTINI2005237}, which can largely save the resources for the identification by measuring a less number of observables~\cite{ISI000270821000002,ISI000288903600019} or without state initialization~\cite{PhysRevLett.102.187203}. For example, only one local observable is required for Hamiltonian identification of a three-spin-chain system~\cite{ISI000300413700002}. 
Moreover, several systematic approaches were also explored for Hamiltonian identification. In~\cite{BONNABEL20091144,6166432}, a quantum observer was designed for estimating the unknown parameters of a qubit, which is driven by the measurement results. Similarly, a Bayesian estimation method was presented for a two-qubit system, which can fulfill the Hamiltonian identification under noisy measurement data~\cite{PhysRevA.80.022333}. In addition, quantum process tomography~\cite{PhysRevA.80.010101} and quantum state tomography~\cite{8022944} approaches were generalized to the Hamiltonian identification, which rely on the accurate estimation of the evolutions or the states in the time-consuming tomography process. Besides the above time-domain approaches, a class of transfer-function-based frequency-domain approaches was presented for spin networks~\cite{PhysRevLett.113.080401}, where the unknown parameters can be obtained by solving a set of nonlinear algebraic equations induced by a measurement-result-induced realization. This method was experimentally realized in a nuclear magnetic resonance system~\cite{HOU2017863}.
However, the performance of the above methods would degrade when they are applied to real quantum systems  which are open in general; i.e., coupled to a thermal bath or other quantum systems resulting in their dissipative dynamics.

For the identification of open quantum systems, several quantum identification methods were presented.
The transfer-function-based method in~\cite{PhysRevLett.113.080401} was extended to Markovian quantum systems~\cite{PhysRevA.91.052121}, i.e., the quantum system in a memoryless environment with a short correlation time. A similar work can also be found in~\cite{ZhouWW}. Moreover, a maximum likelihood estimation method was proposed for an atomic maser system where the detection process can be considered as a discrete-time Markovian chain~\cite{PhysRevA.83.062324, ISI000343895800010}. Alternatively, continuous measurements can be applied for the Hamiltonian identification of finite-level open quantum systems~\cite{INSPEC5492137,6389714,ISI000332614700001}. Moreover, classical identification methods were extended to
linear quantum systems such as quantum harmonic-oscillator networks~\cite{7130587,PhysRevA.95.033825}. 
However, when the quantum system involves interacting both finite-level and infinite-level sub-systems, which is common in cavity QED systems, the underlying complicated algebra and nonlinear dynamics make the identification much harder. To our knowledge, such problems have not been explored in the literature.

In this paper, we propose a gradient algorithm for the identification of such quantum systems. To this aim, we describe the open quantum system by a general Markovian master equation, and the time traces of a selected set of observables are used for identifying the Hamiltonian such that the state of the system at a time instant can be expressed in terms of a series of evolution superoperators acting on the initial state of the system.
We verify the effectiveness of our algorithm with a Markovian Jaynes-Cummings model in cavity (circuit) or quantum electrodynamical systems~\cite{wallraff2004,Mi156}.
We also apply our algorithm to learn the spectrum of a non-Markovian environment associated with a colored quantum noise with a long correlation time~\cite{XuePRA2012,Xue_2017}. In the literature, a differential algorithm~\cite{PhysRevA.82.062104} and a gradient algorithm~\cite{8537939} were designed for the identification of a damping rate function that characterizes the non-Markovian environment in a time-convolution-less master equation. The identification can be also done in the frequency-domain, where the kernel function of a non-Markovian environment is embedded in an integral-differential master equation~\cite{PhysRevA.87.022324}. In our previous work, we have shown that a non-Markovian environment can be modelled by using linear ancillary quantum systems where the spectrum of the environment is parameterized via a spectral factorization theorem. Hence, a non-Markovian quantum system can be modeled in an augmented Hilbert space, whose dynamics are described by a Markovian master equation~\cite{7605518,xue2017arxiv}. In this paper, we assume that the dynamics of the non-Markovian quantum system with the unknown spectrum of the non-Markovian environment obey the augmented  Master equation. The identification of the spectrum is ascribed to estimate the parameters of the ancillary systems in the augmented system model. We apply our gradient algorithm to identify those parameters so as to recover the spectrum of the non-Markovian environment. The feasibility of our algorithm for the identification of the environment is illustrated in an example of a qubit in a non-Markovian environment with a two-Lorentzian spectrum.


This paper is organized as follows. In Section~\ref{sec2}, we describe the open quantum system by a master equation with unknown parameters. Then, a gradient algorithm for identification of the Markovian quantum systems is presented in Section~\ref{sec3}. The performance of our algorithm is verified by an example of a quantum-dot-resonator system in Section~\ref{sec4}. In Section~\ref{sec5}, we show that our algorithm can be applied to explore the spectrum of a non-Markovian environment. Conclusions are drawn in Section~\ref{sec6}.
\section{Description of a Markovian quantum system with unknown parameters}\label{sec2}
A Markovian quantum system is referred to as a quantum system interacting with a memoryless environment~\cite{Breuer}. For example, the dynamics of a cavity mode interacting with a vaccum field is usually Markovian, where the field can be considered as quantum white noise~\cite{Breuer}. This kind of systems widely exists in quantum systems.

In this paper, we describe Markovian quantum systems by a master equation
\begin{eqnarray}\label{1}
  \dot{\rho}(t)=\mathcal{L}\rho(t)=(\mathcal{L}_0+\mathcal{L}_\theta)\rho(t),
\end{eqnarray}
where $\rho(t)$ is the density operator of the Markovian quantum system to be identified. Its evolution is determined by two superoperators $\mathcal{L}_0$ and $\mathcal{L}_\theta$ as shown on the right-hand side of (\ref{1}). The first superoperator
\begin{equation}\label{1-1}
\mathcal{L}_0\rho(t)=-i[H_0,\rho(t)]+\sum_{q=1}^Q\lambda_q\mathcal{D}_{L_q}\rho(t)
\end{equation}
represents the known dynamics of the system where $H_0$ is the Hamiltonian of the system and $[\cdot,\cdot]$ is the commutator for two operators. The second term on the right hand side of (\ref{1-1}) describes the $Q$ channels of dissipative processes with the given damping rate constants $\{\lambda_q,q=1,\cdots,Q\}$. The Lindblad superoperator $ \mathcal{D}_{L_q}$ is calculated as
\begin{equation}\label{2}
  \mathcal{D}_{L_q}\rho(t)=L_q\rho(t)L_q^\dagger-\frac{1}{2}L_q^\dagger L_q\rho(t)-\frac{1}{2}\rho(t)L_q^\dagger L_q,~q=1,\cdots,Q
\end{equation}
with a coupling operator $L_q$ of the system for each dissipative channel with respect to quantum white noise. Hereafter, we have assumed that Plank constant $\hbar=1$.

The second superoperator
\begin{equation}\label{2-1}
\mathcal{L}_\theta\rho(t)=-i[\sum_{m=1}^M\theta_m H_m,\rho(t)]+\sum_{n=1}^N\theta_{M+n}\mathcal{D}_{L_n}\rho(t)
\end{equation}
describes the unknown quantum dynamics. The first part corresponds to the unknown coherent dynamics involving $M$ \textit{unknown} parameters $\{\theta_m, m=1,\cdots,M\}$ associated with $M$ Hamiltonians. The second part represents the unknown incoherent dynamics involving $N$ parameters $\{\theta_{M+n}, n=1,\cdots,N\}$ associated with $N$ Lindbladians. Here, $\mathcal{D}_{L_n}\rho(t)$ is written in the same form as (\ref{2}). The Markovian master equation forms the basis for the identification task in this work.



Since the superoperator $\mathcal{L}$ is time-invariant, the formal solution of the master equation (\ref{1}) is expressed as
\begin{equation}\label{3}
  \rho(t)={\rm exp}{\Big \{}\mathcal{L}(t-t_0){\Big \}}\rho(t_0)
\end{equation}
where $t_0=0$ is the initial time of the evolution and $\rho(t_{0})$ is the initial density operator. Note that in general it is difficult to obtain an analytical expression of (\ref{3}) since it involves to calculate the exponential of the superoperator $\mathcal{L}$. Also, in the following calculation, we let $t_0=0$ for convenience.


\section{A gradient algorithm for Hamiltonian identification of Markovian quantum systems}\label{sec3}
\subsection{Measurement of time trace observables}

To identify the unknown parameters in the Markovian master equation, we measure the time trace of a local observable $O$ of the Markovian quantum system. We assume that we can repeatedly initialize the system to some initial state~\cite{PhysRevA.69.050306} or we have many identical copies of the system. Under this assumption, we measure the observable $O$ with an equal sampling time $\Delta t$; i.e., the time interval from an initial time $0$ to a final time $T$ can be divided into $K=T/\Delta t$ equal intervals. At a sampling time, the density matrix of the system is projected onto the basis of the observable $O$. With the measurement data at every sampling times, we can map the states of the Markovian quantum system to the time trace of the observable $O$ which can be written as
\begin{equation}\label{3-1}
 \hat y=[\hat y_1,\cdots,\hat y_k,\cdots,\hat y_K]^T.
\end{equation}
Here, $\hat y_k$ is the real measured time trace of the observable $O$ at the $k$-th sampling time $t_k=k\Delta t$ with $k=1,\cdots,K$.

Since these real measurement results reflect the dynamics of the density matrix $\rho(t)$ affected by the unknown parameters, one can identify these unknown parameters via sufficiently many measurement observables.

\subsection{Description of the Hamiltonian identification problem for the Markovian quantum system}

Due to the measurement process, the evolution of the Markovian quantum system can be discretized. Since the parameters in the master equation (\ref{1}) are all constants,  the density matrix $\rho(t_k)$ at the time $t_k=k\Delta t$ can be calculated as
\begin{equation}\label{4}
  \rho(t_k)={\mathcal{M}}_k\cdots{\mathcal{M}}_2{\mathcal{M}}_1\rho(0),~~k=1,\cdots,K
\end{equation}
where $\rho(0)$ is the initial density matrix of the system.
The discretized superoperator of $\mathcal{L}$ is written in a matrix exponential form as
\begin{eqnarray}\label{5}
  {\mathcal{M}}_\kappa&=&{\rm exp}{\Big \{}\Delta t(\mathcal{L}_0+\mathcal{L}_\theta){\Big \}},~~\kappa=1,\cdots,k.
\end{eqnarray}

Further, when we guess a set of unknown parameters,
using the equation (\ref{4}),
we can calculate the corresponding time trace observable at the time $t_k=k\Delta t$ as
\begin{eqnarray}\label{6}
  y&=&\left[
        \begin{array}{ccccc}
          y_1, & \cdots, & y_k, &\cdots, & y_K \\
        \end{array}
      \right]^T
\end{eqnarray}
with $y_k={\rm tr}[O\rho(t_k)]$. Although the output $y$ is not the real measurement result, it reflects how the given set of parameters affects the output of the system, which can afford a hint for finding the real parameters.

To measure the distance between the real measurement result $\hat y$ and the calculated result $y$, we define an objective function
\begin{equation}\label{7}
  J=\frac{1}{2}\sum_{k=1}^K(y_k-\hat y_k)^2,
\end{equation}
which is a summation of the square of the differences between the real measurement and calculated results at every sampling times.

Thus, the identification problem considered in this paper can be converted to an optimization problem as follows. Given the real measurement results $\hat y$ (\ref{3-1}), the optimization problem is to find a set of the unknown parameters $\{\theta_1,\cdots,\theta_M, \theta_{M+1},\cdots,\theta_{M+N}\}$ that minimizes the objective function $J$ subject to the evolution (\ref{4}).

\subsection{A gradient algorithm for solving the optimization problem}
To solve the optimization problem, we design a gradient algorithm for revealing the real unknown parameters such that we can minimize the corresponding objective $J$.
The core issue for designing the algorithm is to calculate the gradient of the objective $J$ with respect to the unknown parameters, with which we can search an optimal solution along the gradient descent direction.

For the convenience of the following derivation, we rewrite the unknown parameter set as $\{\theta_p,p=1,\cdots,M+N\}$.
By using the chain rule, the gradient of $J$ with respect to the unknown parameters $\{\theta_p,p=1,\cdots,M+N\}$ can be calculated as
\begin{eqnarray}\label{8}
  \frac{\partial J}{\partial\theta_p}
  &=&\sum_{k=1}^K(y_k-\hat y_k)\langle O\frac{\partial\rho_k}{\partial\theta_p}\rangle.
\end{eqnarray}
Since the unknown parameters have been assumed to be constants in every time intervals, they affect the dynamics in every time intervals. Hence, the gradient of $\rho_k$ with respect to the unknown parameters corresponds to  the superoperators ${\mathcal{M}}_{\kappa}$  from the initial $\kappa=1$ up to $\kappa=k$ time intervals, which can be calculated as
\begin{equation}\label{9-1}
\frac{\partial\rho_k}{\partial\theta_p}=\sum_{\kappa=1}^k{\mathcal{M}}_k\cdots{\mathcal{M}}_{\kappa+1}\frac{\partial{\mathcal{M}}_{\kappa}}{\partial\theta_p}{\mathcal{M}}_{\kappa-1}\cdots{\mathcal{M}}_1\rho(0).
\end{equation}


Further, the gradient of the discretized superoperator ${\mathcal{M}}_{\kappa}$ with respect to the unknown parameters can be approximated as
\begin{equation}\label{10-1}
 \frac{\partial{\mathcal{M}}_{\kappa}}{\partial\theta_p} \approx\left\{\begin{array}{ll}
\Delta t\mathcal{L}_m(\cdot){\mathcal{M}}_{\kappa}(\cdot),& p=1,\cdots,M\\
\Delta t\mathcal{D}_{L_n}(\cdot){\mathcal{M}}_{\kappa}(\cdot),& p=M+1,\cdots,M+N\\
\end{array}\right.
\end{equation}
where the superoperator $\mathcal{L}_m$ is defined as $\mathcal{L}_m(\cdot)=-i[H_m,(\cdot)]$. The superoperators $\mathcal{D}_{L_n}$ and ${\mathcal{M}}_{\kappa}$ can be computed as (\ref{2}) and (\ref{5}), respectively.

By utilizing (\ref{8}), (\ref{9-1}), and (\ref{10-1}), we can update the guessed unknown parameters as
\begin{equation}\label{11-1}
 \theta_{p}\rightarrow
 \theta_{p}-\epsilon\frac{\partial J}{\partial\theta_p}
\end{equation}
where $\epsilon$ is the positive step size. Note that we update the parameters along their descent gradient directions and thus the corresponding updated objective function can be smaller than the original one.

We summarize the gradient identification algorithm for the Markovian quantum system as follows.

Step 1: Choose a local observable $O$, measure the time trace of the observable $\hat y$, initialize the state of the system $\rho(0)$ and the step size $\epsilon$, and guess the initial values of the unknown parameters.

Step 2: Calculate the evolution of the density matrices of the Markovian quantum system from $\rho(1)$ to $\rho(K)$ and the outputs from $y_1$ to $y_K$ with the guessed unknown parameters.

Step 3: Compute the objective $J$ and its gradient with respect to $\{\theta_p\}$ according to (\ref{8}), (\ref{9-1}), and (\ref{10-1}).

Step 4: Update $\theta_p$ using the relation (\ref{11-1}).

Step 5: When a termination condition is satisfied, stop the algorithm; otherwise, go to the Step 2 and start a new iteration.


Importantly, our algorithm is designed for a general Markovian quantum system, where we have not specified the Hamiltonian of the system. Compared to the existing identification methods working for either finite-level quantum systems or infinite-level quantum systems, our algorithm works for a general Markovian quantum systems involving both of them.

\section{Identification of unknown parameters in a quantum-dot-resonator system}\label{sec4}
In this example, we test the performance of our algorithm by a quantum-dot-resonator~\cite{Frey} where the charged quantum dot and the resonator are both dissipative. This system can be described by an open Jaynes-Cummings model which is widely used in circuit or cavity quantum electrodynamical systems~\cite{wallraff2004,Mi156}.

The Hamiltonian of this system can be written as
\begin{equation}\label{13}
  H=\frac{\nu_q}{2}\sigma_z+\nu_0 a^\dagger a+g_d(a^\dagger\sigma_-+a\sigma_+),
\end{equation}
where the first two terms on the right-hand side are the internal Hamiltonian of the quantum dot and the resonator, respectively, and the third term describes the interaction between them. $\sigma_z$ is the z-axis Pauli matrix and $\sigma_-$ and $\sigma_+$ are the ladder operators for the quantum dot. The annihilation and creation operators of the resonator are written as $a$ and $a^\dagger$, respectively.
The Hamiltonian involves three parameters including the splitting frequency of the quantum dot $\nu_q$, the angular frequency of the resonator $\nu_0$, and the coupling strength between them $g_d$.
In addition, the coupling operators of the quantum dot and the resonator for each dissipative channel are $L_1=\sigma_-$ and $L_2=a$, respectively. Hence, the dynamics of this open system can be described by a master equation as
\begin{eqnarray}\label{13-1}
  \dot{\rho}(t) &=&-i[\frac{\nu_q}{2}\sigma_z+\nu_0 a^\dagger a+ g_d(a^\dagger\sigma_-+a\sigma_+),\rho(t)] \nonumber \\
  &&+\gamma_d{\Big(}\sigma_-\rho(t)\sigma_+-\frac{1}{2}\sigma_+ \sigma_-\rho(t)-\frac{1}{2}\rho(t)\sigma_+ \sigma_-{\Big) } \nonumber\\
  && +\gamma_0{\Big(}a\rho(t)a^\dagger-\frac{1}{2}a^\dagger a\rho(t)-\frac{1}{2}\rho(t)a^\dagger a{\Big)}.
\end{eqnarray}

\begin{table}
\squeezetable
\caption{Four guessed initial values for the unknown parameters.}\label{table1}
\begin{tabular}{ccc}
  \hline
   $g_{d0}$ & $\gamma_{d0}~$ & $\nu_{q0}$ \\
  \hline
   $0.6{\rm GHz}$~~ & $0.1\pi{\rm GHz}$~~ & $3 {\rm GHz}$\\
   $0.2{\rm GHz}$~~ & $0.15\pi{\rm GHz}$~~ & $7{\rm GHz}$ \\
   $0.5{\rm GHz}$~~ & $0.05\pi{\rm GHz}$~~ & $5{\rm GHz}$ \\
   $0.4{\rm GHz}$~~ & $0.25\pi{\rm GHz}$~~ & $4{\rm GHz}$ \\
  \hline
\end{tabular}
\end{table}
In this paper, we assume that the splitting frequency of the quantum dot $\nu_q$, the coupling strength $g_d$, and the damping rate $\gamma_d$ are unknown to be identified.
We adopt the parameters for the quantum-dot-resonator system in~\cite{Frey} to simulate the dynamics of the real system. We set $\nu_0=6.775{\rm GHz}$, $\nu_q=6.1814{\rm GHz}$, $g_d=0.3142{\rm GHz}$, $\gamma_d=0.6283{\rm GHz}$, and $\gamma_0=2.6\times 2\pi{\rm MHz}$. With these parameters, we can simulate the real measurement result $\hat y$ where the observable $\sigma_x$ of the quantum dot is measured. To start, we randomly choose initial guesses on the unknown parameters given in Table \ref{table1}. Both the step sizes for updating the three parameters are $0.0002{\rm GHz}$ and the termination condition is that the number of the iterations of the algorithm achieves 40000 times. We assume that the quantum dot and the resonator are initialized on the state $\frac{1}{2}(I+\sigma_x)$ and the ground state, respectively. Note that we have truncated the infinite levels of the resonator to twenty levels to obtain the real measurement results in the simulation and eight levels in the identification model, respectively. The identification result is further examined by truncating the oscillator to twenty levels to make sure the eight-level truncation is sufficient. Simulation results show that they are consistent. If not, one can gradually increase the number of levels and use the obtained identification as an initial guess to refine the optimization.

\begin{figure}
  \flushleft
  \includegraphics[width=9cm]{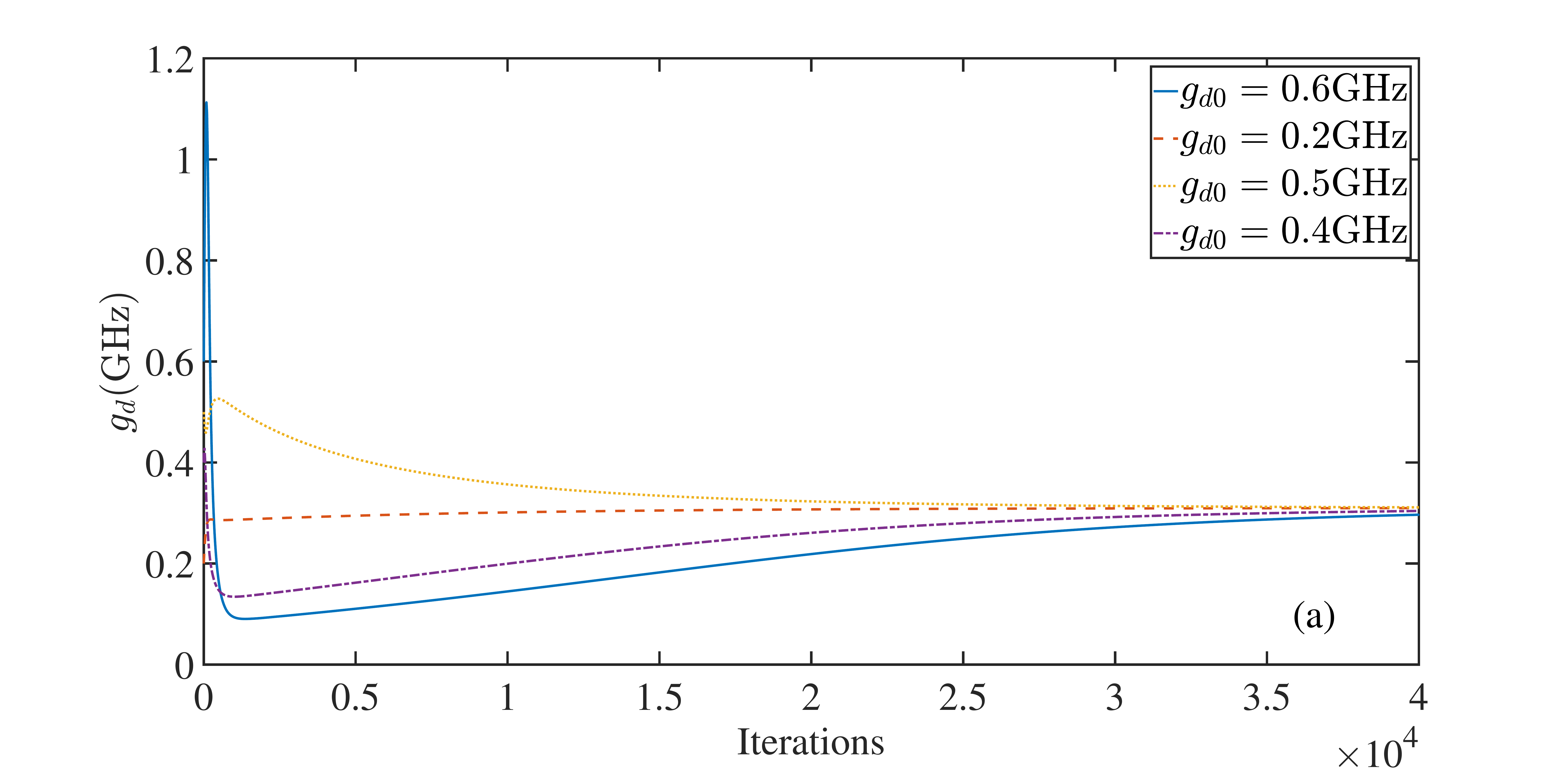}\\
  \includegraphics[width=9cm]{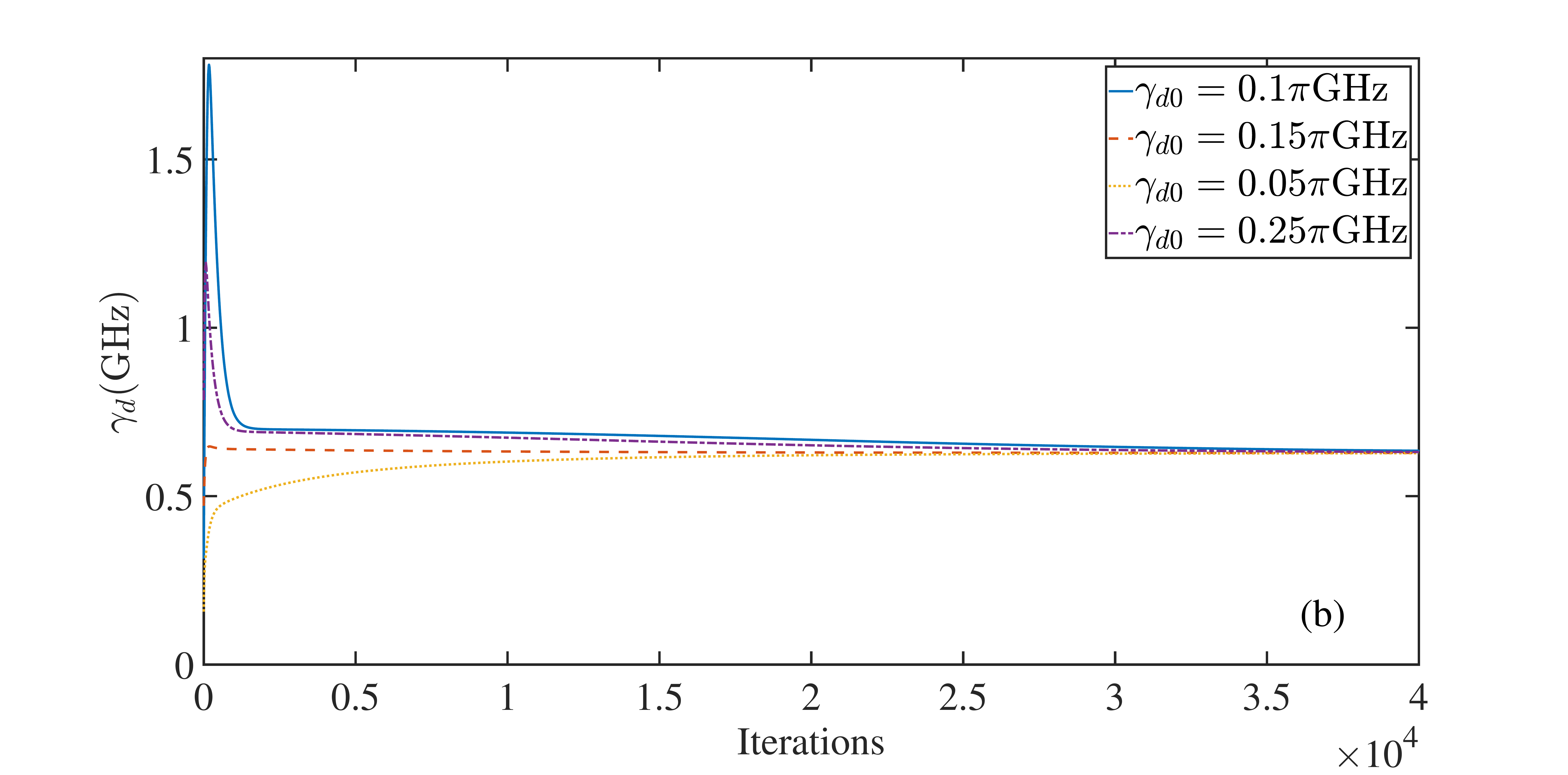}\\
  \includegraphics[width=9cm]{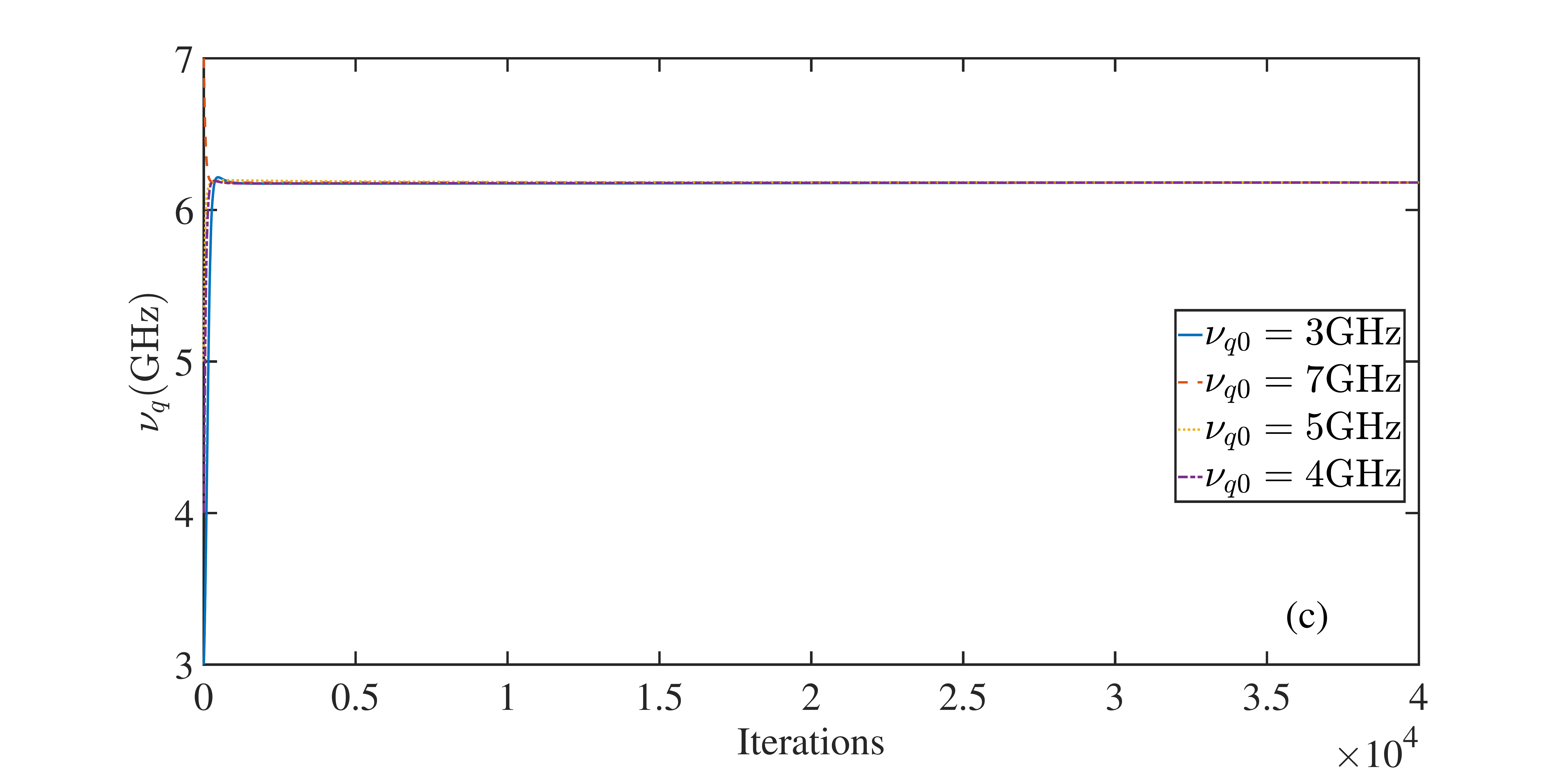}
  \caption{The convergence processes for the identified unknown parameters (a) the coupling strength $g_d$, (b) the damping rate $\gamma_d$ and (c) the splitting frequency $\nu_d$ with four sets of the initial guessed values. The three parameters in the four cases all approach to their real values by using our gradient identification algorithm}\label{figure1}
\end{figure}
\begin{figure}
  \flushleft
  \includegraphics[width=9cm]{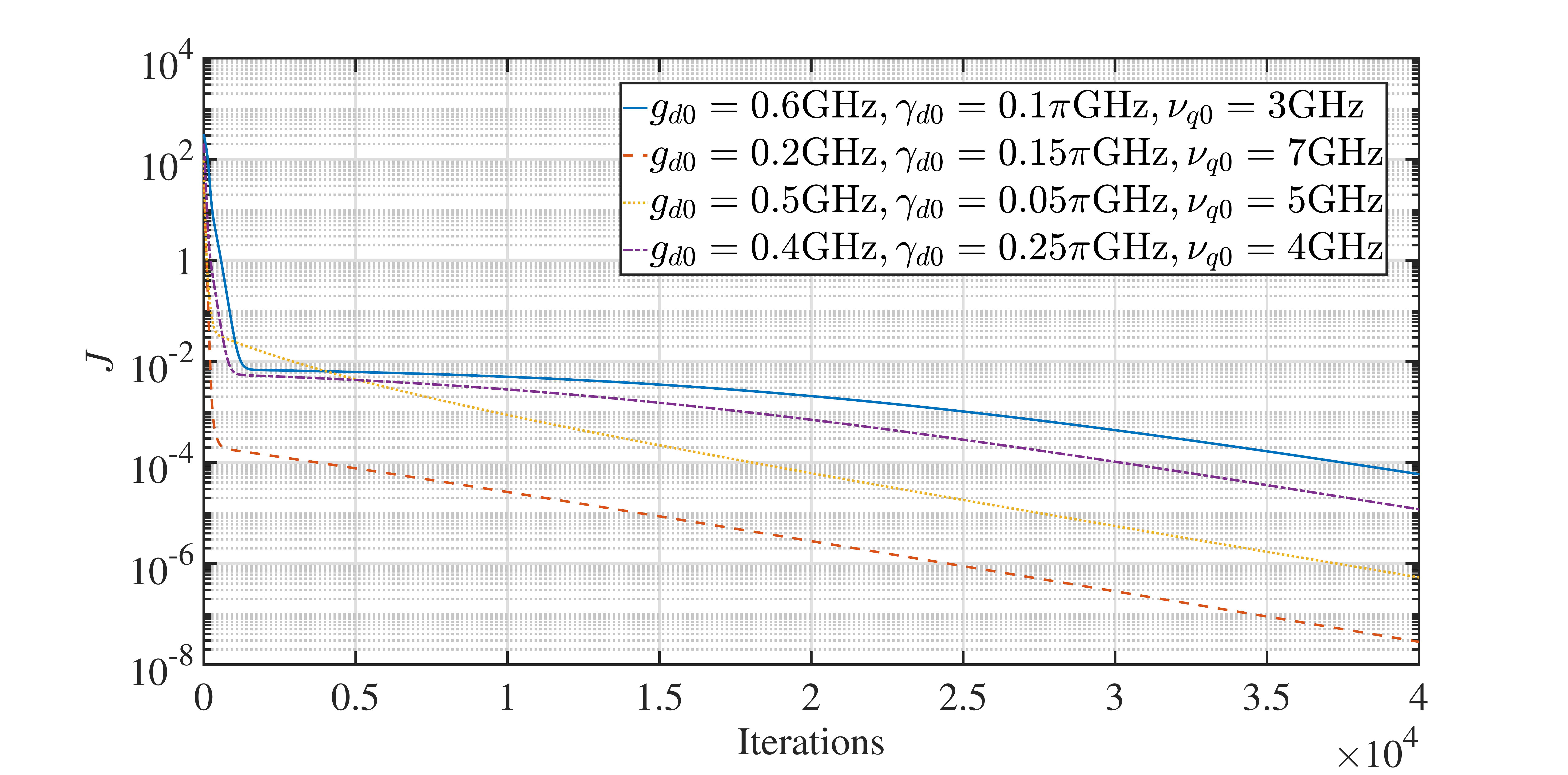}\\
  \caption{The convergence of the objective $J$ in the cases of four initial guessed values of the unknown parameters for the quantum-dot-resonator system. With the suitable initial guess, our gradient identification algorithm can achieve a high precision for the identification.}\label{figure1-1}
\end{figure}

The identification results are given in Fig.~\ref{figure1}. In the cases of four initial guesses, the values of the unknown parameters approach to their real ones, where the converging process of the splitting frequency $\nu_q$ is very fast compared with the slow ones for the other two parameters.
 It shows that our algorithm can iteratively learn the real values of the unknown parameters by utilizing the real measurement results on the expectation values of the observable $\sigma_x$ as shown in Fig.~\ref{figure1}. The variation of the corresponding objective function $J$ during the identification process is shown in Fig.~\ref{figure1-1}. As the identified values of the unknown parameters approaching to the real values, the objective functions $J$ decease monotonically. The objective $J$ is reduced down to $10^{-7}$ with the second initial guess which is lower than all other cases. The best identified values for the three parameters are $\hat g_0= 0.3097{\rm GHz}$, $\hat\gamma_d= 0.6285{\rm GHz}$, and $\hat \nu_q=6.1814{\rm GHz}$. These results indicates that our identification algorithm can achieve a high precision for the identification.

The above simulation shows that our algorithm can identify the unknown parameters in a Markovian quantum-dot-resonator system with a high precision by using only the measurement results of a local observable. Although the algorithm involves an iterative searching process costing computational times, it is acceptable since this process is offline.
\section{Identification of a non-Markovian environment for a qubit system}\label{sec5}
\subsection{An augmented Markovian system model of a non-Markovian qubit}

In solid-state systems, a quantum information carrier involving complicated interactions would exhibit non-Markovian dynamics~\cite{PhysRevB.79.125317}.
The complicated interactions can be explained as the influence of a non-Markovian environment which is characterized by a noise spectrum. To acquire the exact information of the non-Markovian environment, one feasible way is to identify an environment model from the measurement data of the corresponding system, which can obtain a suitable model for the aim of control.

\begin{figure}
  \centering
  \includegraphics[width=8cm]{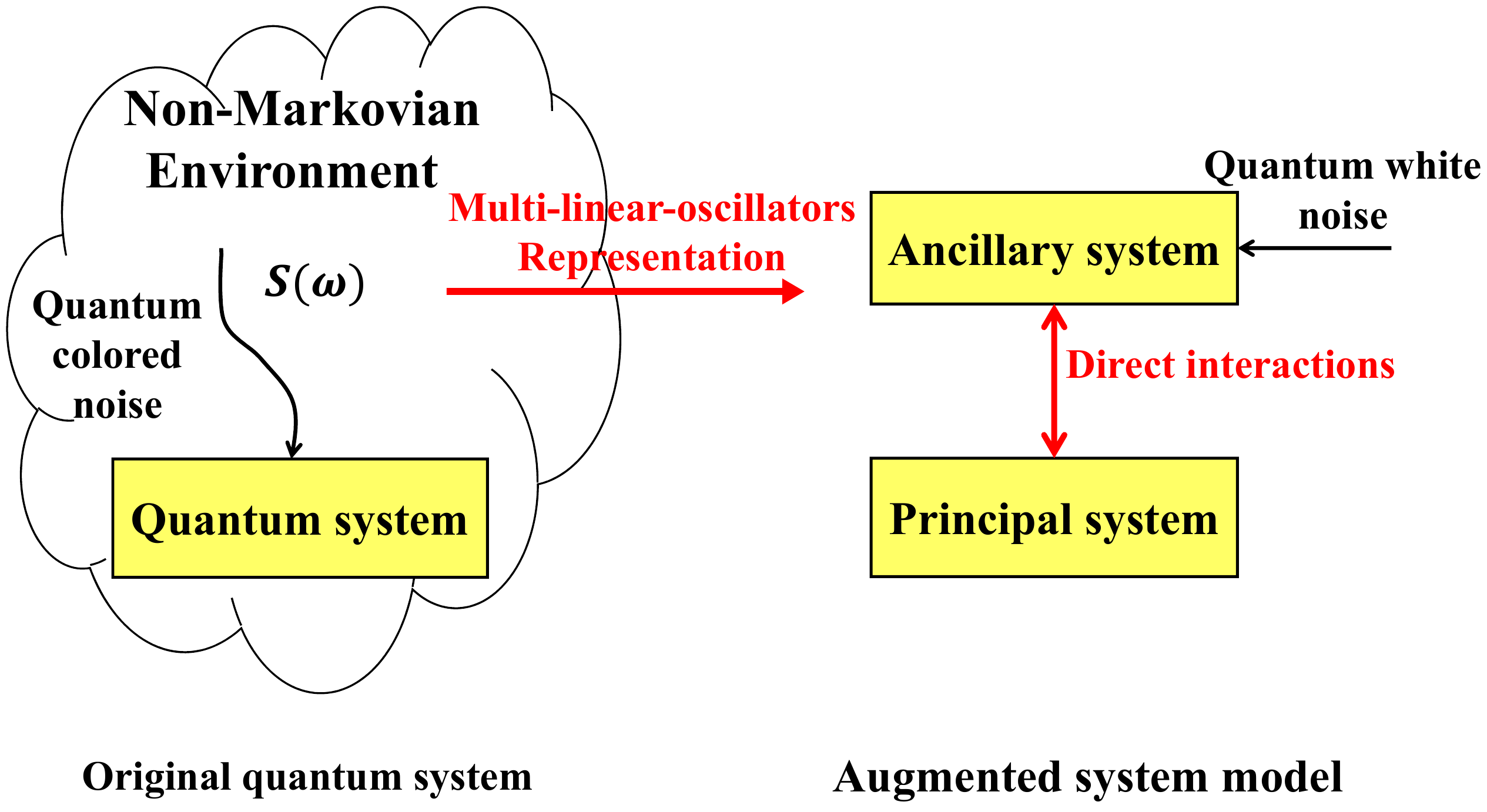}\\
  \caption{A schematic plot of the augmented system model for a non-Markovian quantum system. The original quantum system and the non-Markovian environment are represented by a principal system and an ancillary system, respectively, in the augmented system model.}\label{OQSA}
\end{figure}

In our previous works~\cite{7605518,xue2017arxiv},  an augmented system model approach was presented to describe a non-Markovian quantum system with a given spectrum, whose schematic plot is given in Fig.~\ref{OQSA}.
In the augmented system model, a principal system and an ancillary system represent the original quantum system and the non-Markovian environment, respectively. The ancillary system are several one-mode quantum harmonic oscillators driven by quantum white noise where a one-mode quantum harmonic oscillators can generate quantum Lorentzian noise~\cite{7605518}.
The center frequency, the width, and the strength of the Lorentzian spectrum are determined by the angular frequency, the damping rate with respect to the quantum white noise, and the coupling strength to the principal system of the quantum harmonic oscillator, respectively. Coupling these ancillary systems to an identical operator of the principal system via their direct interactions, the combination of the Lorentzian spectra can approximate noise with an arbitrary spectrum. Here, the direct interactions can capture the mutual influence between the non-Markovian environment and the system. Hence, due to the influence of the ancillary system, the principal system can exhibit non-Markovian dynamics. For more details, see~\cite{7605518,xue2017arxiv} for references.

In this section, we couple a single qubit system to a non-Markovian environment such that the results of the measurement on the qubit can help to access the information of the environment. For the purpose of identification, we assume that the non-Markovian quantum system is described by an augmented system model where the information of the environment corresponds to the parameters of the ancillary system.

Since the quantum colored noise is whitened by the ancillary system, the augmented system is driven by quantum white noise such that the dynamics of the augmented system can be described by a Markovian master equation~\cite{7605518,xue2017arxiv}
\begin{equation}\label{16}
  \dot\rho(t)=(\mathcal{L}_q+\mathcal{L}_i+\mathcal{L}_d)\rho(t),
\end{equation}
where $\rho(t)$ is the density matrix of the augmented system. The superoperator
\begin{equation}\label{17}
  \mathcal{L}_q\rho(t)=-i[H_q,\rho(t)]
\end{equation}
describes the internal dynamics of the principal system. The Hamiltonian of the qubit is $H_q=\frac{1}{2}\omega_0\sigma_z$ with a working frequency $\omega_0$. Moreover, we represent the environment by using many one-mode oscillators so that the internal dynamics of the ancillary systems and the dynamics induced by their couplings to the principal system are described by the superoperator
\begin{equation}\label{17}
\mathcal{L}_i\rho(t)=-i[\sum_{r=1}^R\omega_r^a a_r^\dagger a_r+\sum_{r=1}^Ri\mu_r(a_r^\dagger z_*-z_*^\dagger a_r),\rho(t)]
\end{equation}
where the operators $a_r$ and $a_r^\dagger$ and the frequency $\omega_r$ are the annihilation and creation operators and the angular frequency for the $r$-th ancillary system, respectively. The annihilation and creation operators satisfy the canonical commutation relation~\cite{Breuer}. We have defined the fictitious output of each ancillary system as $c_r=-\frac{\sqrt{{\bar \gamma}_r}}{2}a_r$ which is coupled to the principal system through an operator $\sqrt{\beta_r}z_*$ via their direct interaction $i\mu_r(a_r^\dagger z_*-z_*^\dagger a_r)$ with a coupling strength $\mu_r=-\frac{\sqrt{{\bar \gamma}_r\beta_r}}{2}$. Here, $\bar\gamma_r$ is the damping rate of the $r$-th ancillary system with respect to quantum white noise
 and $\beta_r$ is the coupling strength between the principal system and the $r$-th ancillary system.
 It has been shown that the fictitious output $c_r$ carries a channel of quantum Lorentzian noise. Since they have been coupled to the same operator of the principal system $z_*$, we can combine these Lorentzian noises for generating quantum colored noise with one arbitrary spectrum.
 In addition, the dissipative processes of the ancillary systems with respect to quantum white noise are described by the superoperator
\begin{equation}\label{18}
\mathcal{L}_d\rho(t)=\sum_{r=1}^R\bar\gamma_r\mathcal{D}_{L_r}\rho(t),
\end{equation}
where $L_r=a_r$ is the coupling operator of the $r$-th ancillary system. The number of the ancillary systems $R$ is finite and countable.

Note that with this augmented system model, we can represent a non-Markovian quantum system in a high dimensional Hilbert space as a Markovian quantum system, where the augmented system is only driven by quantum white noise. In the Heisenberg picture, our augmented system model can reproduce the traditional integral-differential Langevin equation~\cite{7605518}. In this sense, the augmented system model is consistent with the traditional model. Also, the quantum colored noise arising in the non-Markovian environment is whitened, whose noise spectrum is parameterized by the ancillary system. Hence, this augmented system model form the bases for identification of the non-Markovian environment.
\subsection{Identification of the non-Markovian environment of a qubit}
\begin{figure}
  \centering
  \includegraphics[width=8.5cm]{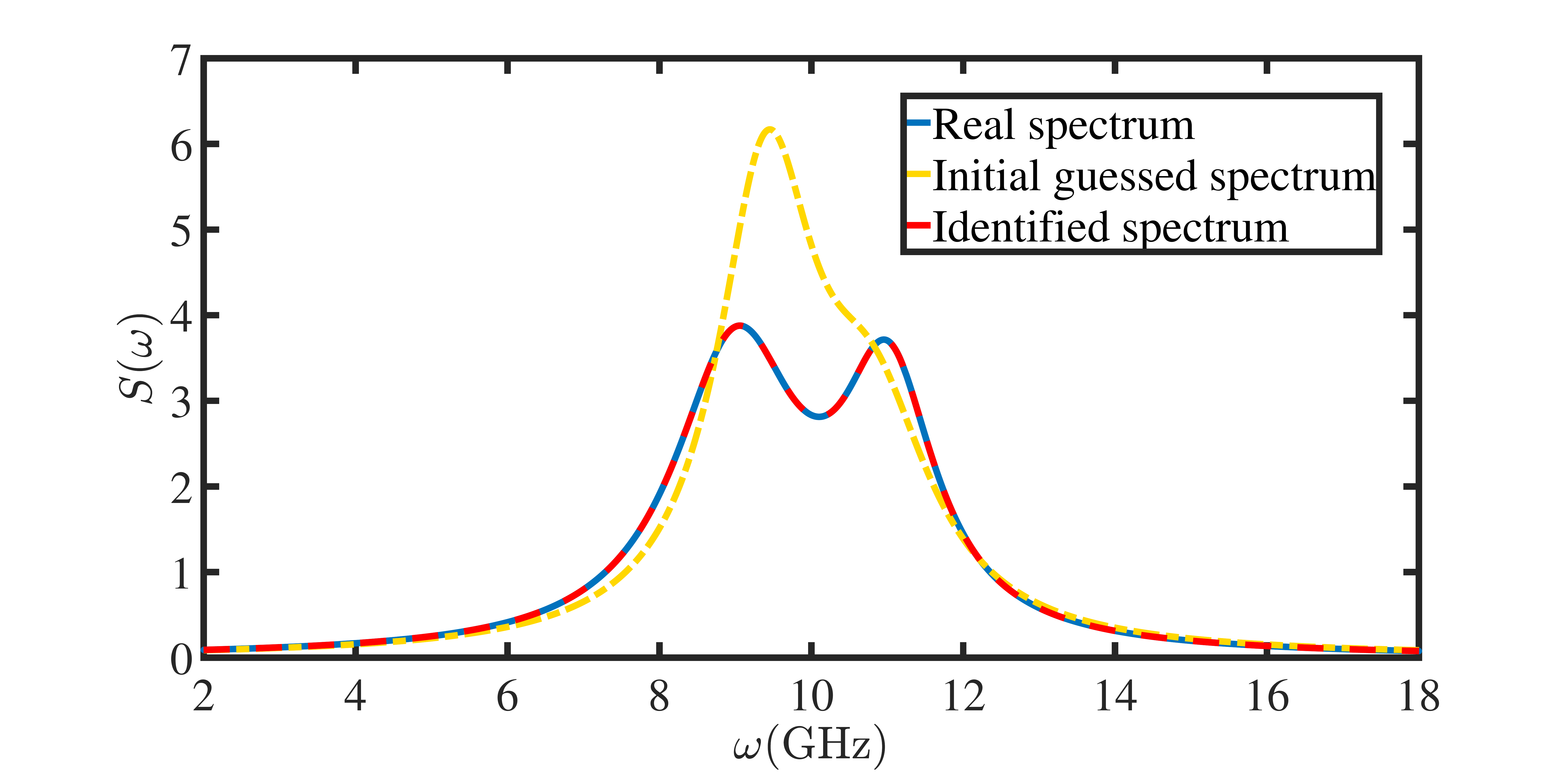}\\
  \caption{The real, initial guessed and identified two-Lorentzian spectra which are plotted as the blue solid, yellow dot-dashed, and red dashed lines, respectively. The identified spectrum obtained by using our gradient algorithm matches the real one very well.}\label{noisespec}
\end{figure}

We consider that the real noise spectrum $S(\omega)$ of the non-Markovian environment for a qubit system is in a two-Lorentzian shape~\cite{Breuer} as
\begin{equation}\label{19}
  S(\omega)=\frac{\beta_1(\frac{\bar\gamma_1}{2})^2}{(\frac{\bar\gamma_1}{2})^2+(\omega-\omega^a_1)^2}+\frac{\beta_2(\frac{\bar\gamma_2}{2})^2}{(\frac{\bar\gamma_2}{2})^2+(\omega-\omega^a_2)^2}.
\end{equation}
The center frequencies of the two-Lorentzian spectrum are $\omega_1^a=9{\rm GHz}$ and $\omega_2^a=11{\rm GHz}$, respectively. The widths of the spectrum are $\bar\gamma_1=2{\rm GHz}$ and $\bar\gamma_2=1.5{\rm GHz}$ and their strengthes are determined by $\beta_1=3.5{\rm GHz}$ and $\beta_2=3{\rm GHz}$, respectively. This spectrum plotted as the blue line in Fig.~\ref{noisespec} is used to characterize the real non-Markovian environment which induces the non-Markovian dynamics of a qubit with an angular frequency $\omega_0=10{\rm GHz}$.
We initialize the qubit in the state $\frac{1}{2}(I+\sigma_x)$. In our simulation, we assume that the ancillary systems are initialized in the ground states. Note that we also truncate the infinite levels of the ancillary linear quantum system to eight levels.
For this system, we also measure the local observable $\sigma_x$ of the qubit and thus we can obtain the real measurement result.

\begin{figure}
  \centering
  \includegraphics[width=6cm]{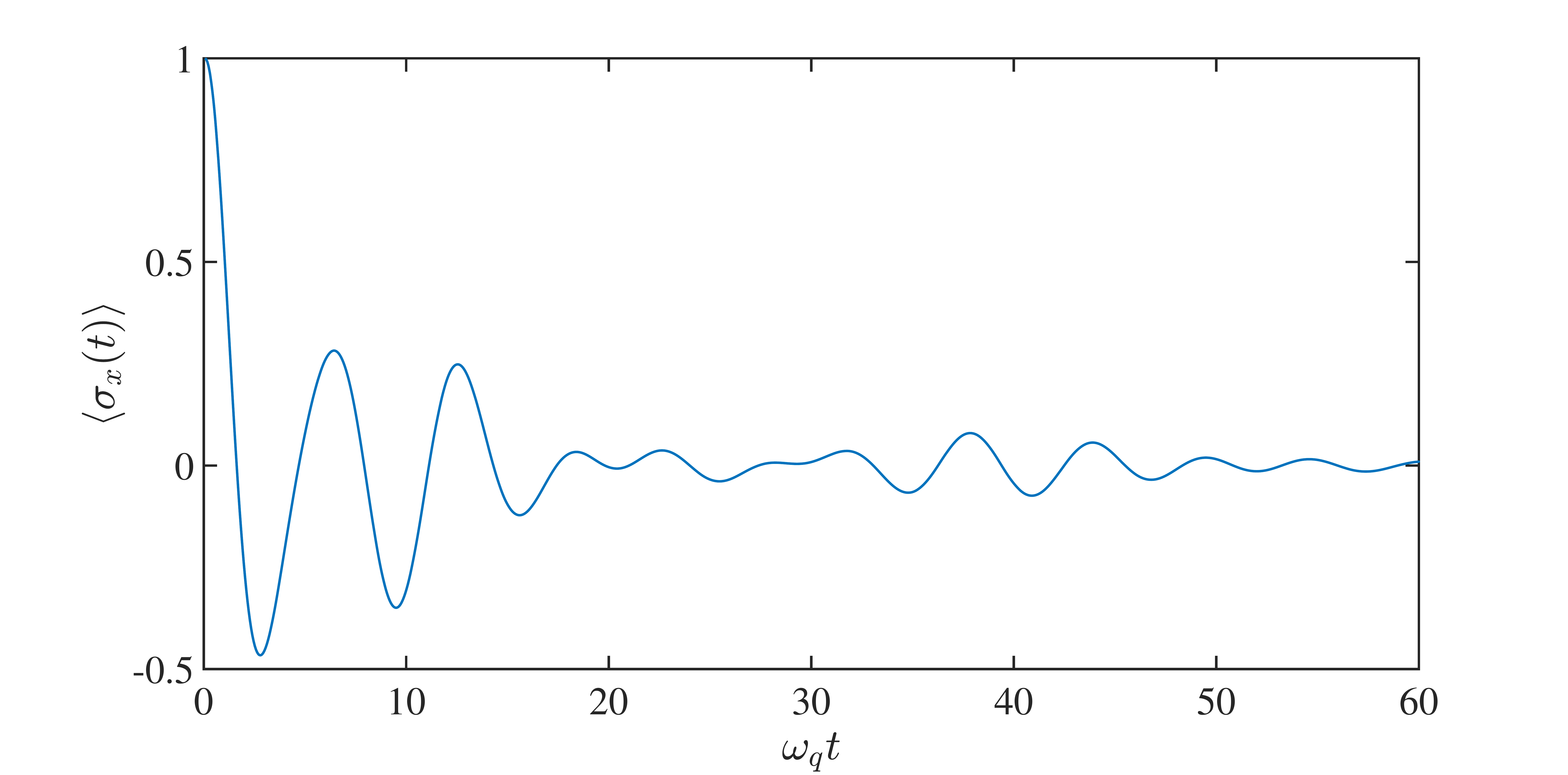}\includegraphics[width=7cm]{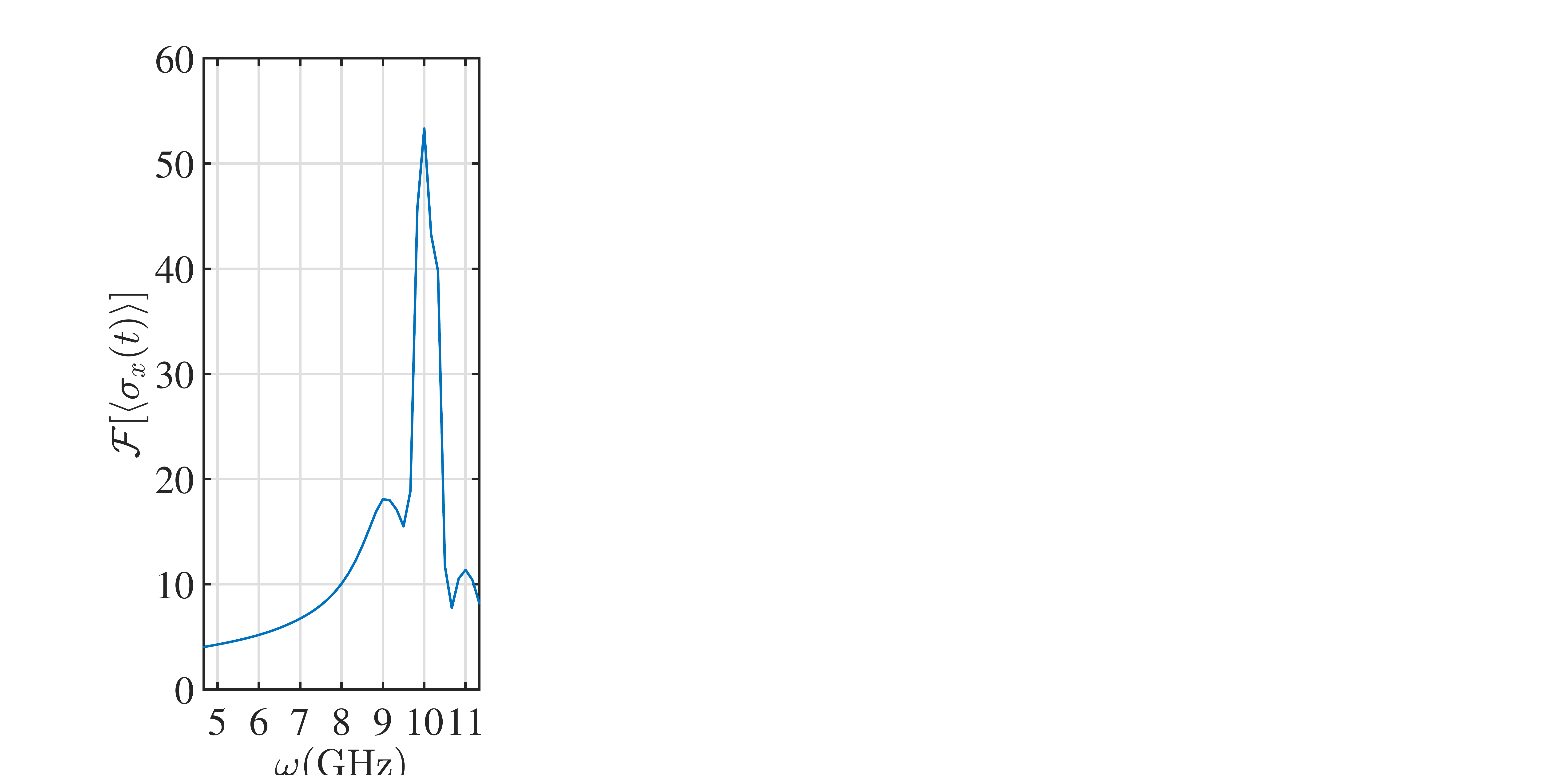}
  \caption{The evolution of the expectation value of the observable $\sigma_x$ and its corresponding Fourier transform. The Fourier transform of the output indicates the number of the ancillary system and help to guess the initial values of the frequency of the ancillary system.}\label{sigmax}
\end{figure}

Before running the gradient identification algorithm, it needs to determine the number of the ancillary systems and guess the initial values of the unknown parameters. These information can be roughly obtained from the measurement result. To the aim, we plot the evolution of the expectation of $\sigma_x$ and its corresponding discrete Fourier transform in Fig.~\ref{sigmax}(a) and (b), respectively. Since the discrete Fourier transform is obtained based on the limited measurement data of the observable $\sigma_x$, the accuracy of the curve for the discrete Fourier transform may not be guaranteed. However, in Fig.~\ref{sigmax} (b), the curve shows that there are two additional modes beside the qubit mode, which indicates that two linear ancillary systems should be coupled to the qubit in the augmented system model. The two additional peaks are around $9{\rm GHz}$ and $11{\rm GHz}$ such that we guess the angular frequencies of the two ancillary systems are $\omega_1^0=9.4{\rm GHz}$ and $\omega_2^0=10.8{\rm GHz}$, where the superscript $0$ represents their initial values. In addition, we estimate that the damping rates of the two ancillary system with respect to quantum white noise are $\bar\gamma_1^0=1.56{\rm GHz} $ and $\bar\gamma_2^0=1.92{\rm GHz}$, respectively, which are difficult to be obtained from Fig.~\ref{sigmax} (b). Note that in the interaction Hamiltonian the coupling strengthes can be redefined as $\mu_1=-\sqrt{\beta_1\bar\gamma_1}/2$ and $\mu_2=-\sqrt{\beta_2\bar\gamma_2}/2$ and thus the interaction Hamiltonian can be written as $i\mu_1(a_1^\dagger\sigma_--\sigma_+a_1)$ and $i\mu_2(a_2^\dagger\sigma_--\sigma_+a_2)$. Hence, the task for identification of the parameters $\beta_1$ and $\beta_2$ can be converted to that for $\mu_1$ and $\mu_2$, respectively, whose guessed values are $\mu_1^0=-1.45{\rm GHz}$ and $\mu_2^0=-1.07{\rm GHz}$. Hence, in the identification process, we can utilize the values of $\mu_1$ and $\mu_2$ to calculate $\beta_1$ and $\beta_2$. The step sizes for updating these parameters in our algorithm are all $0.002{\rm GHz}$.

\begin{figure*}
  \centering
  \includegraphics[width=6.5cm]{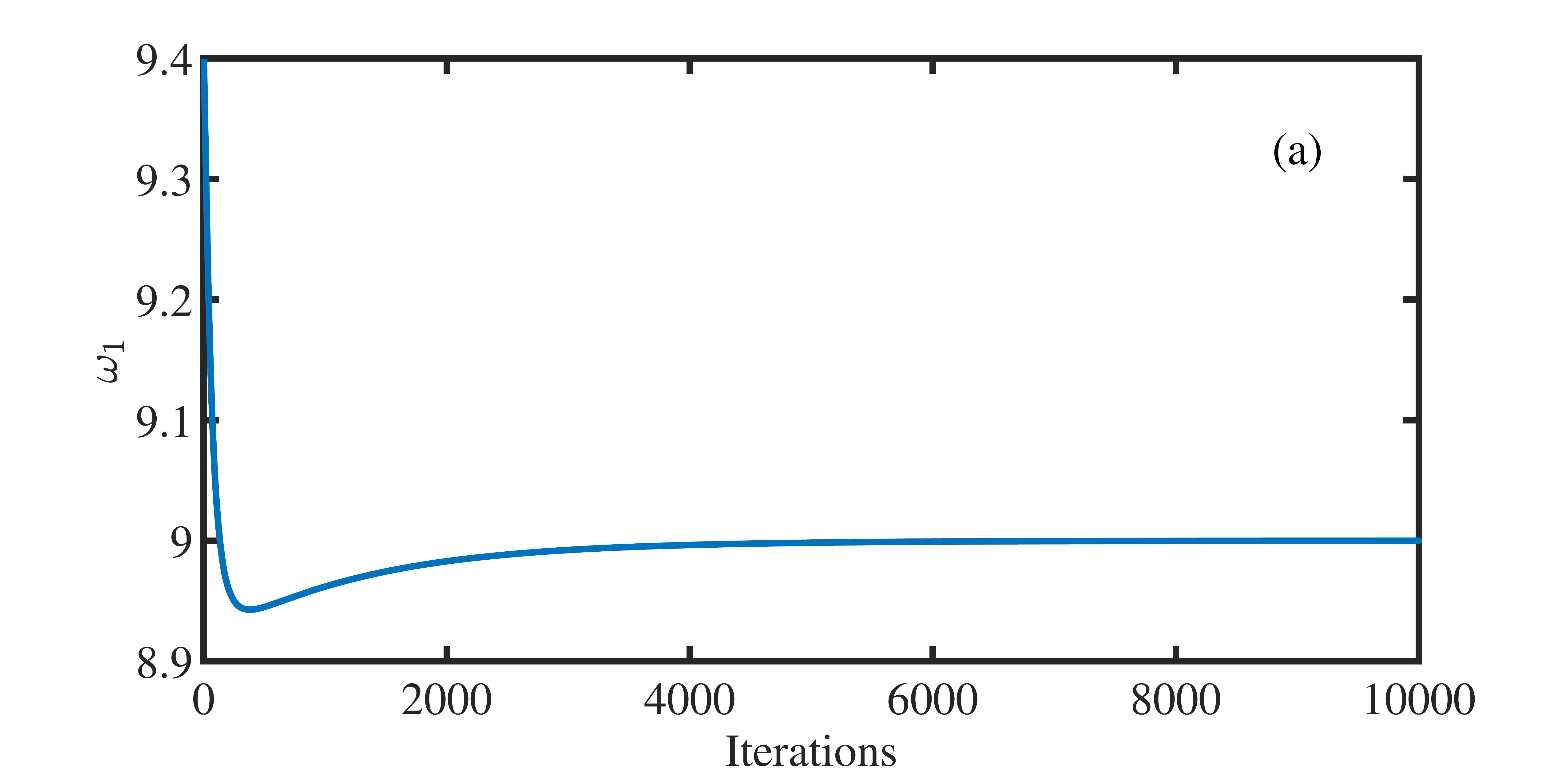}\includegraphics[width=6.5cm]{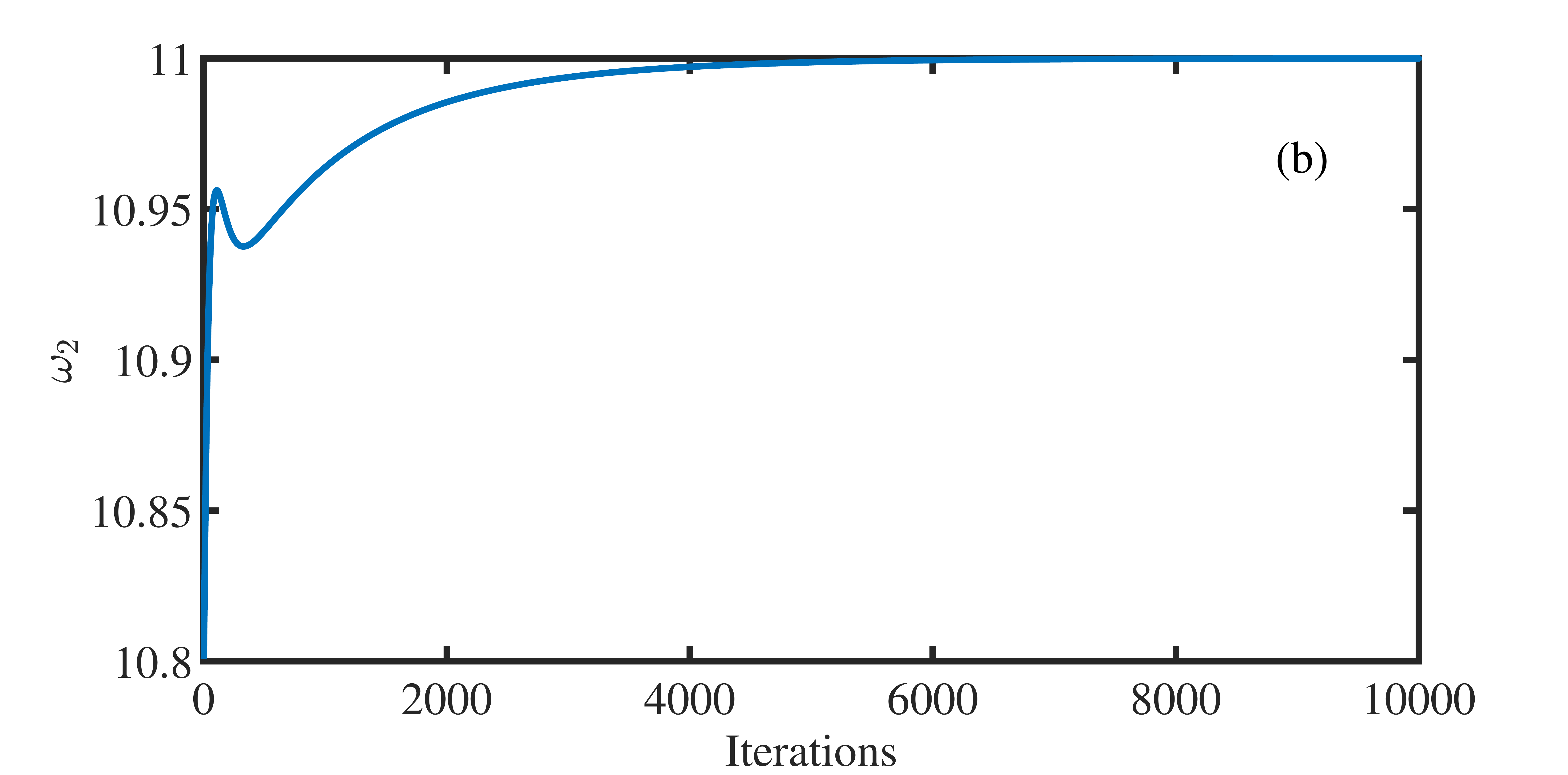}\\
  \includegraphics[width=6.5cm]{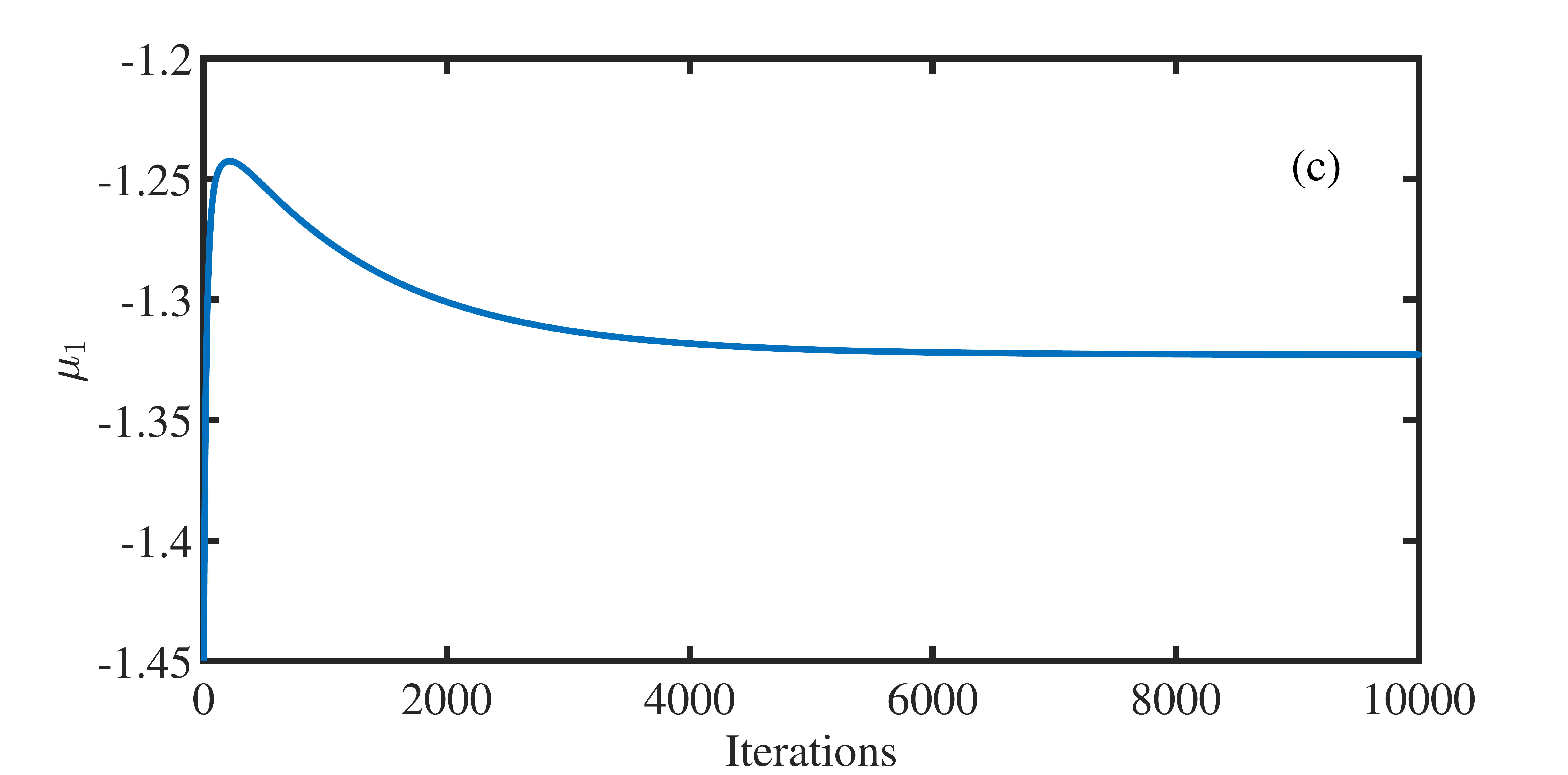}\includegraphics[width=6.5cm]{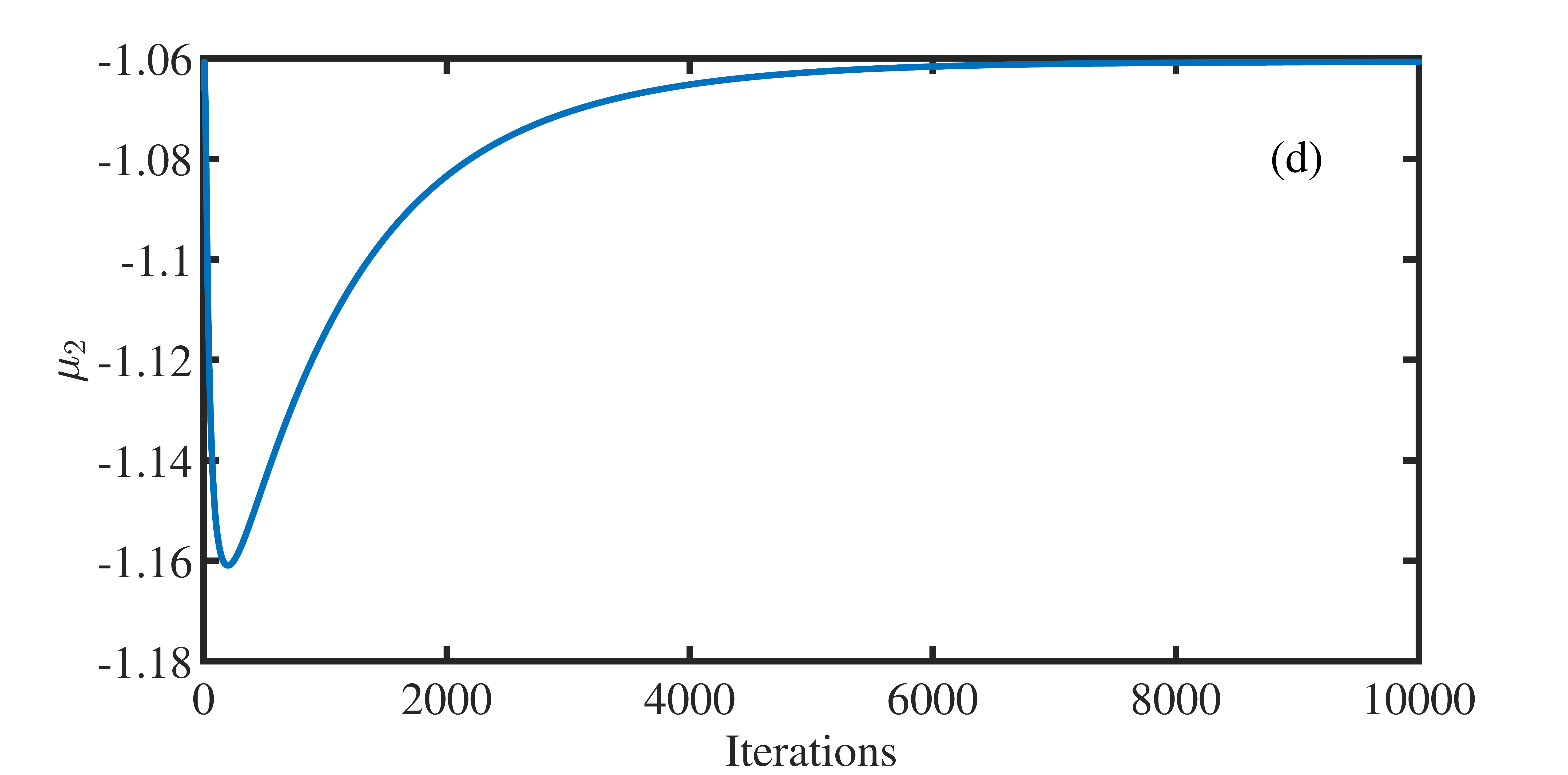}\\
  \includegraphics[width=6.5cm]{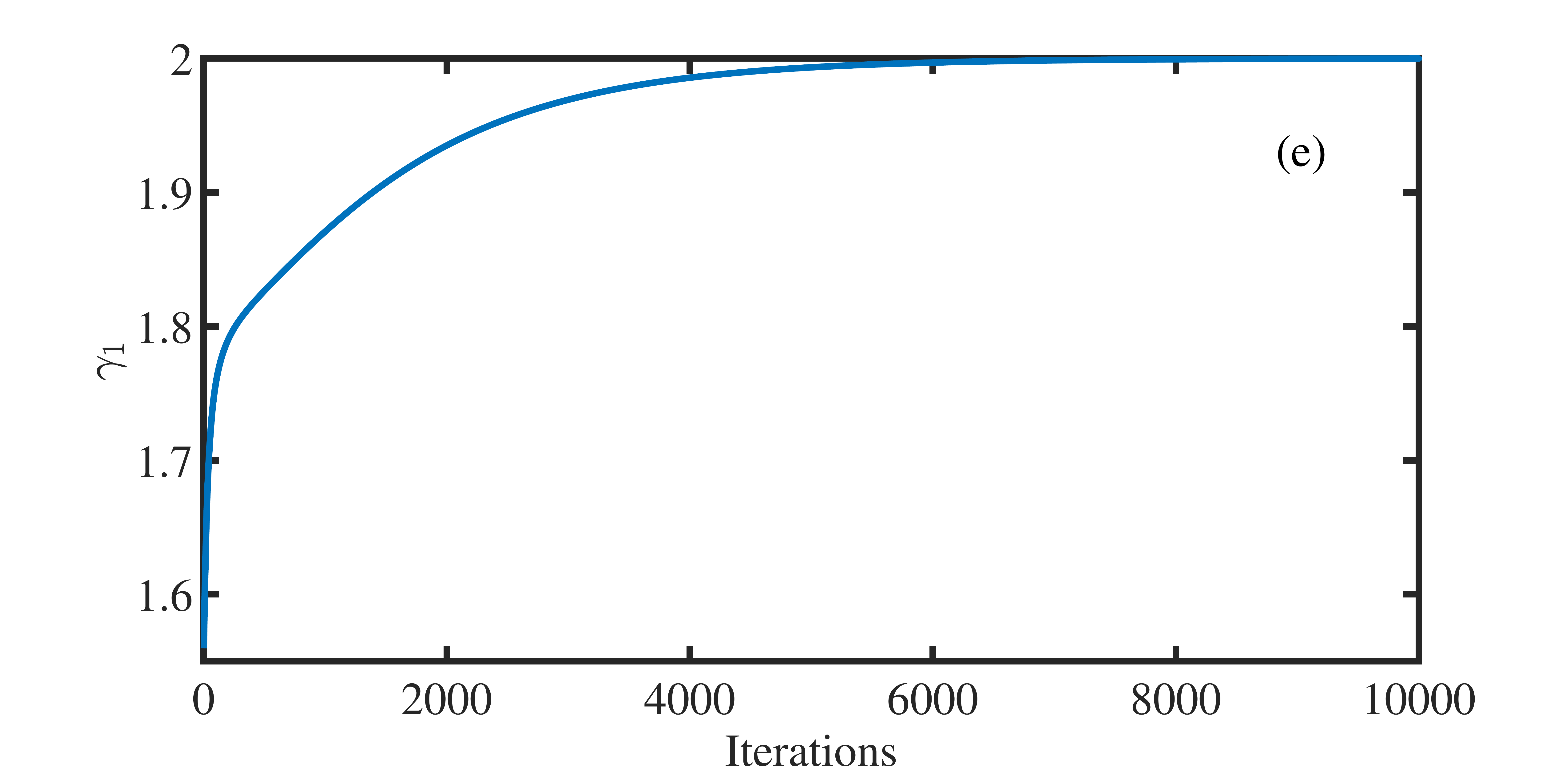}\includegraphics[width=6.5cm]{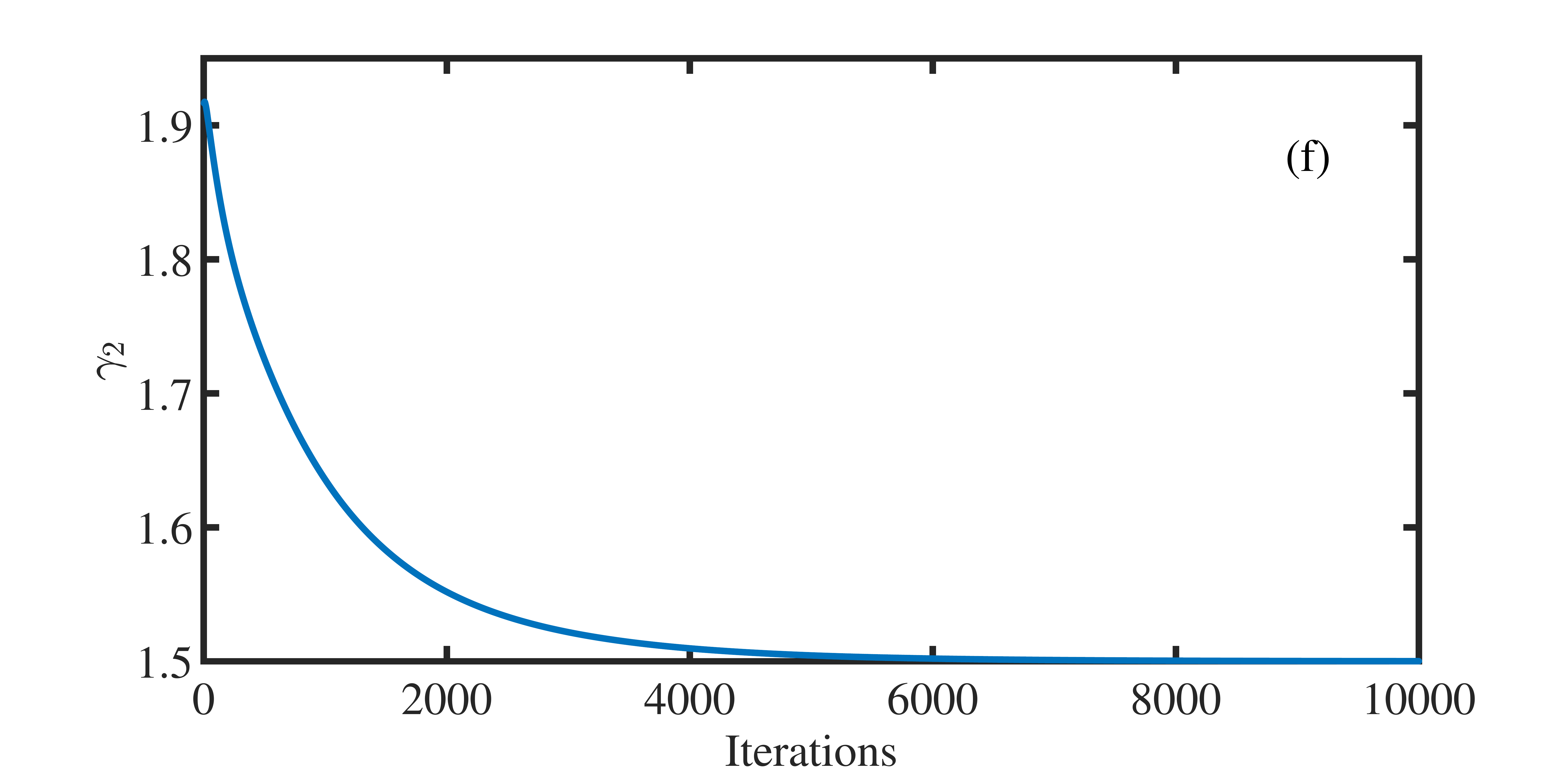}
  \caption{Variations of the identified parameters (a) $\omega_1$, (b) $\omega_2$, (c) $\mu_1$, (d) $\mu_2$, (e) $\gamma_1$ and (f) $\gamma_2$ of the ancillary systems in the augmented system model. The final results are close to the real values very well.}\label{params6}
\end{figure*}

\begin{figure}
  \centering
  \includegraphics[width=9cm]{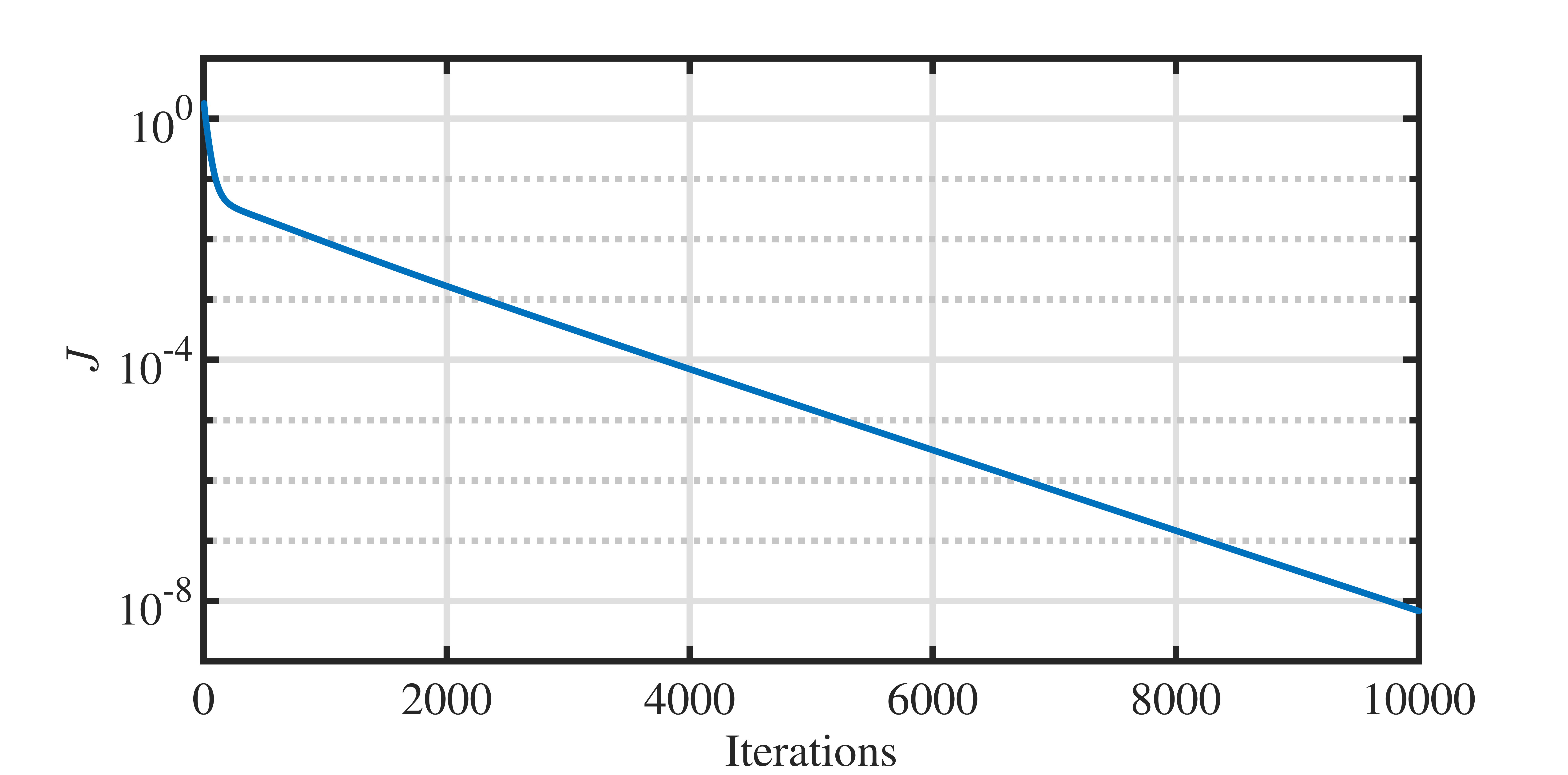}\\
  \caption{Variation of the objective $J$ when we identify the unknown parameters for the non-Markovian environment.}\label{obj2}
\end{figure}

We run the identification algorithm for 10000 iterations and the identification results are given in Fig.~\ref{params6} and Fig.~\ref{obj2}. The final identified frequencies and the damping rate constants are $\omega_1^f=9.0000{\rm GHz}$, $\omega_2^f=11.0000{\rm GHz}$ and $\gamma_1^f=1.9999{\rm GHz}$, $\gamma_2^f=1.5001{\rm GHz}$. From the results for $\mu_1$ and $\mu_2$, we can obtain $\beta_1^f=3.4998{\rm GHz}$ and $\beta_2^f=3.0000{\rm GHz}$.
These identification results are quite close to the real values. During the iterations, the objective function $J$ is reduced down to about $10^{-8}$, which is shown in Fig.~\ref{obj2}. With respect to these identified parameters, we plot the identified spectrum in Fig.~\ref{noisespec} as the red dashed line, which matches the real spectrum very well.
\section{Conclusion}\label{sec6}
In this paper, we have designed a gradient algorithm for the identification of unknown parameters in open quantum systems. This algorithm can utilize the measured data of the time trace observables of the open quantum systems to iteratively learn the real values of the unknown parameters with high accuracies. 
We verified the performance of the gradient algorithm in the example of the quantum-dot-resonator system and applied it to identifying the non-Markovian environment based on the augmented system model. This algorithm  works for a more general Markovian quantum systems, which are important in the calibration of parameters of devices in an experiment with a high accuracy or to explore the interactions between a quantum system and its environment.
\section*{Acknowledgment}
This work was supported
in part by the National Natural Science Foundation of China under
Grants 61873162 and 61473199, in part by the Shanghai Pujiang Program under Grant 18PJ1405500, in part by the Suzhou Key Industry Technology Innovation
Project SYG201808, in part by the Key Laboratory of
System Control and Information Processing in Ministry of Education of China Scip201804, and in part by the Open Research Project of the State Key Laboratory of Industrial Control Technology, Zhejiang University, China No.ICT1900304.

\end{document}